\documentclass[twocolumn,showpacs,preprintnumbers,amsmath,amssymb]{revtex4}

\usepackage{amsmath,amssymb}
\usepackage[latin1]{inputenc}
\usepackage{dcolumn}
\usepackage{bm}

\newcommand{\im}[0]{\text{Im}\,}
\newcommand{\re}[0]{\text{Re}\,}
\usepackage{graphicx}
\usepackage{epsfig}
\usepackage{dcolumn}
\usepackage{bm}

\begin{document}


\title{On the electromagnetic properties of active media}

\author{Bertil Nistad and Johannes Skaar}
\affiliation{Department of Electronics and Telecommunications\\
Norwegian University of Science and Technology}

\date{\today}

\begin{abstract}
Several results concerning active media or metamaterials are proved and discussed. In particular, we consider the permittivity, permeability, wave vector, and refractive index, and discuss stability, refraction, gain, and fundamental limitations resulting from causality.
\end{abstract}

\pacs{41.20.Jb, 42.25.Bs}
\maketitle

\section{Introduction}
The ability of fabricating artificial materials with sophisticated electromagnetic properties have generated large interest recently. By tailoring the electric permittivity and the magnetic permeability, one can obtain arbitrary metamaterial-induced coordinate transformations, leading to devices such as perfect lenses and electromagnetic cloaks \cite{pendry2000,pendry2006,leonhardt2006}. Unfortunately, the performance of these devices are strongly limited by loss. To overcome this limitation it has been suggested to introduce gain \cite{ramakrishna_prb,noginov06,popov06,shalaev_nphot}. 

It is crucial to understand fundamentally how active media behave electromagnetically. Active media are not new; lasers and optical amplifiers are well-established technologies. The new aspects are rather the flexibility of metamaterials and their applications. For lasers, gain saturation plays a vital role when modeling the medium. Nevertheless, in applications such as optical amplifiers, the medium behaves linearly; thus the linear susceptibilities and the complex refractive index characterize its operation. For potential applications to metamaterials, perfect lenses, manipulating the near-fields, cloaks etc., linear operation is also desired. Nevertheless, gain saturation may for example limit the resolution associated with the perfect lens, so nonlinearity needs in many cases to be taken account of.

Here we will limit the discussion to linear media. In other words, the fields are nowhere allowed to be larger than the threshold for gain saturation. Provided there are no instabilities, this may be achieved by ensuring that the excitations are sufficiently weak. (The fields are however considered classical such that the stimulated emission dominates the spontaneous emision.) On the other hand, if there are instabilities, a linear, frequency-domain model is certainly no longer valid. However, if we assume that the medium is dark at some time $t=0$ (for example by having the power supply turned off at $t<0$), a linear transient analysis is still valid until the fields have grown above the gain saturation threshold \footnote{For an accurate description of the transient behavior of unstable systems, perturbations associated with (initiation of) the pump should strictly be included into the analysis.}. This is the standard approach to analyze active, possibly unstable systems in other contexts such as electronics and control engineering. In the present work, the transient analysis is central for understanding the properties of a semi-infinite medium (Section \ref{sign}).

Recently there has been some confusion about how to determine the sign of the refractive index (or the direction of the wave vector) in active media \cite{mackay:159701, ramakrishna:059701,ramakrishna_ol,grigorenko_ol,nazarov07,govyadinov:191103,Boardman:07}. The confusion in the field is probably due to a nice property of passive media: One does not have to invoke causality in its most primitive form to determine the sign of the refractive index. One can determine its sign straightforwardly at a single frequency by requiring that the Poynting vector points away from the excitation, or by requiring that the wave decays as it propagates away from the excitation. This makes it natural to consider general methods for identifying the sign from the (relative) permittivity $\epsilon(\omega)$ and permeability $\mu(\omega)$ at a single frequency $\omega$, even for active media. While such methods may give correct answer for a restricted class of media, they are necessarily incorrect in general. Indeed, there exist both positively and negatively refracting, nonmagnetic media with $\epsilon(\omega)=1-2i\alpha$, $0<\alpha\ll 1$, at a single frequency \cite{skaar06}: Conventional gain media refracts positively, while the right-handed negative index medium suggested by Chen et. al. \cite{chen05}, refracts negatively. To find the sign of the refractive index, we must go back to first principles, using causality in its most primitive form: The front velocity of an electromagnetic wave cannot travel faster than $c$, the vacuum velocity of light. (Actually, the correct solution can be found without invoking causality as an extra principle, as causality is built into the Maxwell equations.) Similarly to the classical treatment of passive media \cite{brillouin}, this amounts to requiring the refractive index to be analytic in some upper half-plane of complex frequency, and requiring $n\to +1$ as $\omega\to\infty$ \cite{skaar06}. However, one must watch out for absolute instabilities; the half-plane may be $\im\omega>\gamma>0$ instead of $\im\omega>0$.
 
The purpose of the present article is five-fold: First we will consider microscopic causality of active media (Section \ref{causality}). As opposed to passive media, it turns out that causality is not sufficient for establishing Kramers--Kronig relations. This has to do with possible instabilities. Second, we will clarify how to resolve the refractive index and the direction of wave vector, and pay particular attention to the ambiguity mentioned above (Sections \ref{sign} and \ref{3d}): Two media with identical $\epsilon(\omega)$ and $\mu(\omega)$ at $\omega$ may respond completely different to monochromatic excitations at $\omega$. In particular we analyze a slab of thickness $d$ in 2D. Assuming stability, a nonmagnetic medium with $\epsilon(\omega)=1-2i\alpha$ at a single frequency ($0<\alpha\ll 1$) will refract positively when $d$ is small, and negatively when $d$ is large. In other words, all media that make the slab stable for small $d$ refract positively at the interfaces, while all media that make the slab stable for large $d$ refract negatively. Third, we discuss and categorize the different instabilities that may arise in active media (Section \ref{instabilities}), and relate them to previous literature in plasma physics. Fourth, we will prove that for active media, any refractive index function in a finite bandwidth can be realized approximately, with any precision (Section \ref{limitations}). Fifth, we consider the ultimate limits of active media, and show that in general, there are no upper limit of the gain associated with media without absolute instabilities (Section \ref{boundsgain}). Also we show that there are no lower bound for the (maximal) gain associated with right-handed, negative index media. Permittivity functions of such media with arbitrarily low gain are provided, giving directions to practical realizations.

\section{Causality for active media}\label{causality}
As mentioned above, we restrict ourselves to linear media. Furthermore, the media are assumed isotropic, homogeneous, and without spatial dispersion. We also assume that the medium is time-shift-invariant, that is, if an excitation is shifted in time, the response shifts by an equal time. Consider the case where the applied electric field is due to a external source with charge density $\rho_e(t)$ \cite{dolgov81}. The effect of the source is to induce a charge density $\rho_i(t)$, which together with $\rho_e(t)$ makes up the total charge density $\rho(t)=\rho_e(t)+\rho_i(t)$. Since the medium is linear and shift-invariant, we have the following relation between the source and the induced charge density:
\begin{equation}\label{causalcharges}
\rho_i(t)=\int_0^\infty x_e(\tau)\rho_e(t-\tau) d\tau.
\end{equation}
Here $x_e(t)$ is a real response function, and the lower limit 0 in the integral is due to causality; the induced charge density cannot precede the source. We will assume that the induced charges do not blow up faster than exponentially for any bounded excitation, which can be formulated as follows:
\begin{equation}\label{blowup}
\int_0^t|x_e(\tau)|d\tau<C\exp(\gamma t),
\end{equation}
for some nonnegative constants $C$ and $\gamma$. Transforming \eqref{causalcharges} to the frequency domain, we can write $\rho_i(\omega)=x_e(\omega)\rho_e(\omega)$ and therefore
\begin{equation}\label{causalchargesomega}
\rho(\omega)=[1+x_e(\omega)]\rho_e(\omega),
\end{equation}
in an obvious notation. Note that for active media in general, the transformation must be performed with the Laplace transform, since the induced charge density may blow up with time. This amounts to requiring all quantities to vanish for negative time, and putting $\im\omega\geq\gamma$ in the integrals. For example,
\begin{equation}\label{laplacerhoe} 
\rho_e(\omega)=\int_0^\infty \mathcal \rho_e(t)\exp(i\omega t)dt.
\end{equation}
Eq. \eqref{causalchargesomega} together with the Maxwell equations $\epsilon_0\nabla\cdot{\bf E}=\rho$ and $\nabla\cdot{\bf D}=\rho_e$, and the constitutive relation ${\bf D}=\epsilon\epsilon_0 {\bf E}$, mean that
\begin{equation}
\epsilon(\omega)=\frac{1}{1+x_e(\omega)},
\end{equation}
or
\begin{equation}
\frac{1}{\epsilon(\omega)}=1+\int_0^\infty x_e(t)\exp(i\omega t)dt.\label{permittivitycaus}
\end{equation}
Since $x_e(t)$ is real we also have the symmetry
\begin{equation}
\epsilon(-\omega^*)=\epsilon^*(\omega)\label{permittivitysym}.
\end{equation}
We note that it is not the permittivity itself, but its reciprocal, that is causal in the sense of being the transform of a response function that vanishes for $t<0$ \cite{dolgov81}.

At very high frequencies, the electrons behave essentially as if they were free, yielding the asymptotic form \cite{nussenzveig}
\begin{equation}\label{asympt}
|\epsilon(\omega)-1|\sim \frac{1}{\omega^2}, \quad\omega\to\infty.
\end{equation}

While causality \eqref{permittivitycaus}, symmetry \eqref{permittivitysym}, and the asymptotic form \eqref{asympt} are valid in general, Kramers--Kronig relations cannot be established unless more information on the media is known. In other words, even though media that satisfy the ususal Kramers--Kronig relations are causal, not all causal media satisfy the Kramers--Kronig relations. This is particularly true for active media, which may exhibit certain instabilities. For example, the response function $x_e(t)$ may increase exponentially, which leads to a singularity of $1/\epsilon(\omega)$ in the upper half-plane $\im\omega>0$. Or the transformed response function $x_e(\omega)$ may be equal to $-1$ somewhere in the upper half-plane, giving a singularity of $\epsilon(\omega)$ there. 

In light of Titchmarsh' theorem \cite{nussenzveig,titchmarsh}, if no such singularities are present, not even at the real frequency axis, \eqref{permittivitycaus} and \eqref{asympt} give the Kramers--Kronig relations
\begin{subequations}
\label{KK}
\begin{align}
&\im \epsilon(\omega)=\frac{2\omega}{\pi}\mathcal P\int_{0}^{\infty}\frac{\re\epsilon(\omega')-1}{\omega^2-\omega'^2}d\omega',\label{KK1}\\
&\re \epsilon(\omega)-1=\frac{2}{\pi}\mathcal P\int_0^\infty\frac{\im\epsilon(\omega')\omega'}{\omega'^2-\omega^2}d\omega',\label{KK2}
\end{align}
\end{subequations}
where $\mathcal P$ denotes the Cauchy principal value \footnote{Eq. \eqref{asympt} ensures that $\epsilon(\omega)-1$ and $1/\epsilon(\omega)-1$ are square integrable. Eq. \eqref{permittivitycaus} then shows that $1/\epsilon(\omega)-1$ is analytic and uniformly square integrable along any line parallell to the real axis, in the upper half-plane. In the absence of singularities of $\epsilon(\omega)$, the analyticity and uniformly square integrability for $1/\epsilon(\omega)-1$ translate into identical properties for $\epsilon(\omega)-1$. Thus the real and imaginary parts of $\epsilon(\omega)-1$ form a Hilbert transform pair.}. If $\epsilon(\omega)$ has singularities at the real axis, the Kramers--Kronig relations must be modified. For example, if the medium is conducting at zero frequency, $\epsilon(\omega)$ is singular at $\omega=0$. Then, the Kramers--Kronig relations are retained if we subtract the singularity, i.e., make the substitution $\epsilon(\omega)\to\epsilon(\omega)-i\sigma/\omega$ in \eqref{KK}, where $\sigma>0$ is the zero frequency conductivity \cite{landau_lifshitz_edcm}. If $\epsilon(\omega)$ have singularities in the upper half-plane, $\epsilon(\omega)$ loses its meaning at real frequencies and the Kramers--Kronig relations cannot be expressed along the real axis. The observation frequency $\omega$ and the integral must instead be taken along a line above the singularities. We will discuss media with instabilities further in Section \ref{instabilities}.

We can treat the magnetic permability in a similar fashion, yielding
\begin{equation}\label{permeabilitycaus}
\mu(\omega)=1+\int_0^\infty x_m(t)\exp(i\omega t)dt,
\end{equation}
instead of \eqref{permittivitycaus}. Here $x_m(t)$ is a real response function. With a similar asymptotic form as \eqref{asympt}, and in the absence of singularities in the upper half-plane, we may also obtain Kramers--Kronig relations for the magnetic permeability. Using a Kramers--Kronig relation analogously to \eqref{KK2} to calculate $\re\mu(0)$ from $\im\mu(\omega)$, we straightforwardly find that $\re\mu(0)>1$ whenever $\im\mu(\omega)>0$. This is not always true \cite{martin67}, as demonstrated by the existence of passive, diamagnetic media. In other words, the Kramers--Kronig relations for the magnetic permeability must be treated with care.

For the remaining parts of this paper, we will only assume causality in the sense \eqref{permittivitycaus} and \eqref{permeabilitycaus}, meaning that we allow for instabilities. However, to retain the meaning of complex frequencies, we exclude superexponential instabilities; in other words \eqref{blowup} and a similar bound for $x_m(t)$ are required \footnote{We are not aware of any existing media with super-exponential instabilities.}. Similarly to \eqref{asympt} we will assume 
\begin{equation}\label{epsilonmuasymptrealomega}
\epsilon(\omega), \mu(\omega)\to 1 \text{ as }\omega\to+\infty,
\end{equation}
whenever real frequencies are meaningful, and Eqs. \eqref{permittivitycaus} and \eqref{permeabilitycaus} imply
\begin{equation}\label{epsilonmuasympt}
\epsilon(\omega), \mu(\omega)\to 1 \text{ as }\im\omega\to+\infty.
\end{equation}

\section{Direction of the wave vector and sign of the refractive index}\label{sign}
To identify the direction of the wave vector and the sign of the refractive index in active media, one can use the following approach \cite{skaar06}: By a transient Laplace transform analysis of a slab of thickness $d$, one may compute the fields for time $t<d/c$, where $c$ is the vacuum velocity of light. Then, by causality, the fields have not felt the presence of the far end; thus they must be identical to those of a semi-infinite medium. By subsequently taking the limit $d\to\infty$, one can extract the wave vector or refractive index. Below we will obtain the same results as in \cite{skaar06} more directly, by taking the limit $d\to\infty$ immediately in the region of convergence of the Laplace transformed fields.
\begin{figure}
  \centering
  \includegraphics{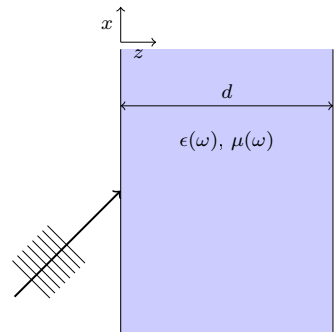}
  \caption{(Color online) Slab of thickness $d$.}
  \label{fig:slab}
\end{figure}
Consider a slab with permittivity $\epsilon(\omega)$ and permeability $\mu(\omega)$, surrounded by vacuum, see Fig. \ref{fig:slab} Since the medium may be active, the fields may blow up with time. Consequently, steady-state frequency-domain fields do not necessarily exist. The natural remedy is to consider Laplace transformed fields instead of nonexisting Fourier transforms, assuming the field is equal to zero for negative time:
\begin{equation}\label{laplace} 
E(\omega)=\int_0^\infty \mathcal E(t)\exp(i\omega t)dt.
\end{equation}
Here $\mathcal E(t)$ denotes the real, physical, time-domain electric field. The Laplace transform \eqref{laplace} exists in a region of convergence $\im\omega>\gamma$, where $\gamma$ is a sufficiently large, positive parameter, such that the exponential factor ensures convergence of the integral. The inverse transform is given by
\begin{equation}\label{invlaplace}
\mathcal E(t)=\frac{1}{2\pi}\int_{i\gamma-\infty}^{i\gamma+\infty}E(\omega)\exp(-i\omega t)d\omega.
\end{equation}

Let the Laplace transformed, incident wave at $z=0^-$ be $E^+(\omega)=E_0^+(\omega)\exp(i k_x x+i k_y y)$, where $E_0^+(\omega)$ is independent on the spatial coordinates, and $k_x$ and $k_y$ are the transversal, spatial frequencies of the source. Solving Maxwell's equations in the transform domain, we obtain the field
\begin{eqnarray}
&& E(\omega)/E^+(\omega) \\
&& = 
\begin{cases}
\exp(ik_zz)+R\exp(-ik_zz) & \text{for $z=0^-$},\\
S^+\exp(ik_z'z)+S^-\exp(-ik_z'z) & \text{for $0\leq z\leq d$},\\
T\exp(ik_z(z-d)) & \text{for $z>d$},
\end{cases} \nonumber
\end{eqnarray}
where the total reflection coefficient $R$, field amplitudes in the slab ($S^+$ and $S^-$), and transmission coefficient $T$ are found to be
\begin{subequations}
\label{slabrt}
\begin{align}
&R=\frac{(\mu^2 k_z^2-{k_z'}^{2})[1-\exp(2ik_z'd)]}{(\mu k_z+k_z')^2-(\mu k_z-k_z')^2\exp(2ik_z'd)}, \label{slabr}\\
&S^+=\frac{2\mu k_z(\mu k_z+k_z')}{(\mu k_z+k_z')^2-(\mu k_z-k_z')^2\exp(2ik_z'd)}, \label{slabsp}\\
&S^-=\frac{2\mu k_z(\mu k_z-k_z')}{(\mu k_z-k_z')^2-(\mu k_z+k_z')^2\exp(-2ik_z'd)},\label{slabsm}\\
&T=\frac{4\mu k_zk_z'\exp(ik_z'd)}{(\mu k_z+k_z')^2-(\mu k_z-k_z')^2\exp(2ik_z'd)}.\label{slabt}
\end{align}
\end{subequations}
Here, $k_z^2=\omega^2/c^2-k_x^2-k_y^2$ and ${k_z'}^{2}=\epsilon\mu\omega^2/c^2-k_x^2-k_y^2$. The expressions above apply to TE polarization; for TM polarization, the expressions are valid provided $\epsilon$ and $\mu$ are interchanged. Note that the fields are invariant if $k_z'\to-k_z'$, so the choice of the sign of $k_z'$ does not matter. 

For active media, $R$, $S$, and $T$ may contain singularities in the upper half-plane. Thus the fields should only be evaluated in their region of convergence $\im\omega > \gamma$, where the line $\omega=i\gamma$, $\gamma>0$ is located above all singularities \cite{papoulis}. Note that singularities in the upper half-plane have nothing to do with noncausality; they rather imply instabilities \cite{skaar06}. In practice, diverging fields imply that the slab starts lasing (taking gain saturation into account).

To obtain the fields associated with a semi-infinite medium, we may take the limit $d\to\infty$. From \eqref{epsilonmuasympt}, it is clear that $k_z'\approx\pm\omega/c$ for sufficiently large $\im\omega$. Since the sign of $k_z'$ is arbitrary in \eqref{slabrt}, we take $k_z'\approx +\omega/c$ there. With this choice, $\im k_z'$ is positive for sufficiently large $\im\omega$. It is not difficult to realize that the singularities of $S$ do not move towards $\omega=i\infty$ as the thickness $d\to\infty$; therefore we may take the limit $d\to\infty$ to find the field in a region $\im\omega>\gamma$. The resulting expression is given by 
\begin{equation}\label{fresnel}
S=\frac{2\mu k_z}{\mu k_z+k_z'}\exp(ik_z'z).
\end{equation}
We have argued for \eqref{fresnel} for $\im\omega>\gamma$. The real, time-domain field is obtained by inverse transforming $S E^+(\omega)$ along the Bromwich path $\omega=i\gamma$. However, by analytic continuation and Cauchy's integral theorem, we may move the integration path down towards the first non-analytic point of \eqref{fresnel}; in other words we may now set $\gamma=\max\{0,\,\im(\text{zero of $\mu k_z+k_z'$}),\,\im(\text{nonanalytic point of $k_z'$})\}$. In many cases of interest, we obtain $\gamma=0$, which means that \eqref{fresnel} and $k_z'$ may be interpreted at real frequencies. In these cases, $k_z'(\omega)$ is the analytic continuation of the branch of $\sqrt{\epsilon\mu\omega^2-k_x^2-k_y^2}$ that tends to $+\omega/c$ as $|\omega|\to\infty$. {\it Thus if possible, the sign of $k_z'$ must be chosen such that $k_z'$ is analytic in the upper half-plane, and such that $k_z'\to+\omega/c$ as $|\omega|\to\infty$. If this is not possible due to poles or odd-order zero(s) of $\epsilon\mu\omega^2-k_x^2-k_y^2$, there are instabilities. Then $k_z'$ does not have any physical significance for real frequencies.} The nature of instabilities associated with poles or odd-order zeros of $\epsilon\mu\omega^2-k_x^2-k_y^2$ will be discussed in the next section. Instabilities associated with zeros of $\mu k_z+k_z'$ are related to the boundary of the medium, and may be eliminated by choosing another surrounding material \cite{skaar06}.

Taking $k_x=k_y=0$, we have $k_z'=n(\omega)\omega/c$, where $n(\omega)=\pm\sqrt{\epsilon(\omega)\mu(\omega)}$. Thus, provided $\epsilon(\omega)\mu(\omega)$ does not have any poles or odd-order zeros in the upper half-plane, $n(\omega)$ is identified as the analytic function in the upper half-plane that tends to $+1$ as $\omega\to\infty$. This result was used in Ref. \cite{chen05} to determine the refractive index in certain active, nonmagnetic, negatively refracting media. When $n(\omega)$ can be identified as an analytic function in the upper half-plane, and $\epsilon(\omega)\mu(\omega)$ is continuous at real frequencies (except possibly at $\omega=0$), $n(\omega)$ must be continuous at real $\omega\neq 0$. Thus we can identify $n(\omega)$ by the simple formula
\begin{equation}
n(\omega)=\sqrt{|\epsilon(\omega)||\mu(\omega)|}\exp[i(\varphi_\epsilon(\omega)+\varphi_\mu(\omega))/2)],
\label{eq:ndef}
\end{equation}
where $\varphi_\epsilon(\omega)+\varphi_\mu(\omega)$ is the complex argument of $\epsilon(\omega)\mu(\omega)$, unwrapped such that it is continuous for $\omega\neq 0$ and such that it tends to $0$ as $\omega\to\pm\infty$. If $\epsilon(\omega)\mu(\omega)$ does have poles and/or odd-order zeros, there will be instabilities, and $n(\omega)$ does not have meaning at real frequencies; only above the nonanalytic points. Note however, that relativistic causality never is violated \cite{skaar06}; $n(\omega)$ is always analytic in {\it some} upper half-plane $\im\omega>\gamma\geq 0$. 
 
Due to the phase unwrapping procedure, the sign of $n(\omega)$ at a particular, real frequency is determined from the global properties of the functions $\epsilon(\omega)$ and $\mu(\omega)$ \cite{chen05,chen06,skaar06,skaar06b}. As demonstrated in Section \ref{3d}, two materials with the same permittivity and permeability at a particular frequency, may have opposite sign of the refractive index. Therefore, {\it any method that identifies the sign of $n(\omega)$ from $\epsilon(\omega)$ and $\mu(\omega)$ at a single frequency $\omega$ must be incorrect in general}. Consequently, although the methods for identifying the sign of $n(\omega)$ in Refs. \cite{mackay:159701,ramakrishna:059701,ramakrishna_ol,grigorenko_ol,nazarov07,govyadinov:191103,Boardman:07} may give the correct result for certain active materials, they are incorrect in general.

A critical point which cannot be overemphasized, is that the complex frequency-domain fields are not the physical fields themselves; they are suitable transforms of the real, physical, time-domain fields (Fourier transforms for passive systems, and Laplace transforms for active systems). Laplace transformed functions exist only in their region of convergence $\im\omega > \gamma$; thus the transformed fields must not be interpreted elsewhere. For example, it is tempting to take $d\to\infty$ while letting $\omega$ be real in \eqref{slabrt} \cite{nazarov07}. However, for media that give rise to singularities of \eqref{slabrt} in the upper half-plane, this is clearly not a correct procedure and may in fact result in incorrect sign for the refractive index. This happens e.g. for a conventional gain medium such as the inverted Lorentzian medium.

Finally, we should keep in mind that any physical medium is limited by gain saturation. When there are no instabilities, the linear model is valid provided the excitations are sufficiently weak. When there are instabilities, the time-domain fields are only valid before they reach the threshold for gain saturation. Provided this happens sufficiently late, transform-domain concepts such as the refractive index and the wave vector have meaning.

\section{Instabilities of causal media}\label{instabilities}
\begin{figure}[ht]
  \centering
  \includegraphics{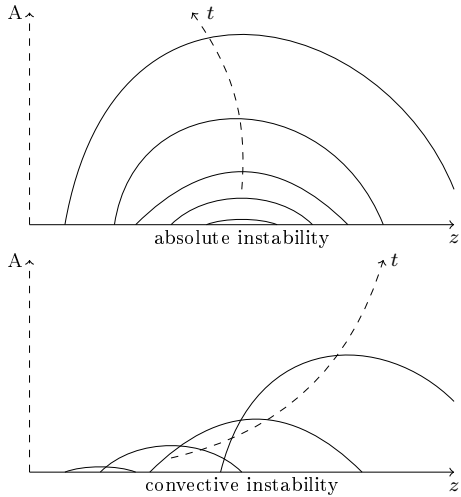}
  \caption{Absolute and convective instabilities.}
  \label{fig:instabilities}
\end{figure}

An instability means that the electromagnetic fields blow up with time. There are three different classes of instabilities \cite{Sturrock,Briggs,Akhiezer}. First we let the medium be infinite. An {\it absolute instability} means that the fields at a fixed point in space blow up with time. A {\it convective instability} means that the fields blow up with time, but at a fixed point, they do not. This means that the instability is ``convected away''. Absolute and convective instabilities are indicated in Fig. \ref{fig:instabilities}. A third category, {\it global instabilities}, arises when the medium is bounded. For example, the fields may diverge due to amplified, multiple reflections.

An example of a convective instability is that of conventional gain media, such as the Erbium-doped fiber amplifier. When such a gain medium is put into a resonator configuration, a global instability may occur. Due to gain saturation (which is a nonlinear process), this leads to lasing rather than infinite fields. 

Absolute and convective instabilities are features of the medium itself, not the total system including boundaries. Absolute instabilities arise when the refractive index $n(\omega)$ cannot be identified as an analytic function in the upper half-plane $\im\omega>0$. This happens if $\epsilon(\omega)\mu(\omega)$ contains poles or odd-order zeros there. 

An apparent absolute instability arises at oblique incidence to a semi-infinite medium \cite{skaar06b}. Even if $n(\omega)$ is analytic in the upper half-plane, $k_z'$ at oblique incidence may have branch points there. This can be interpreted as follows: Any causal excitation contains necessarily an infinite band of frequencies. Some of the frequencies  make $\epsilon\mu\omega^2-k_x^2-k_y^2$ zero; this corresponds to waves propagating perpendicularly to the $z$-axis. Such waves propagates an infinite distance, and therefore pick up an infinite amount of gain, before arriving the plane $z=\text{const}$. This is called a {\it virtual absolute instability}, as it appears mathematically as an absolute instability, while its physical interpretation is similarly to convective instabilities. Virtual absolute instabilities can be eliminated by limiting the extent of the active medium in the transverse direction. 

\section{Stability and refraction in 2D}\label{3d}
We will now limit the discussion to absolutely stable media for which $\epsilon(\omega)$ and $\mu(\omega)$ do not have singularities or zeros in the upper half-plane. (We do not, however, exclude the possibilities of convective or global instabilities.)

The uniqueness theorem for solutions to Maxwell's equations in the frequency domain \cite{pozar} assumes that the medium is lossy. Therefore, it is not surprising that for active media, $\epsilon$ and $\mu$ at a single frequency $\omega$ do not determine whether the material refracts positively or negatively at $\omega$. Indeed, as pointed out earlier \cite{skaar06}, there are both positively and negatively refracting media with identical electromagnetic parameters at a single frequency. 

\begin{figure}[ht]
  \centering
  \includegraphics[width=5cm]{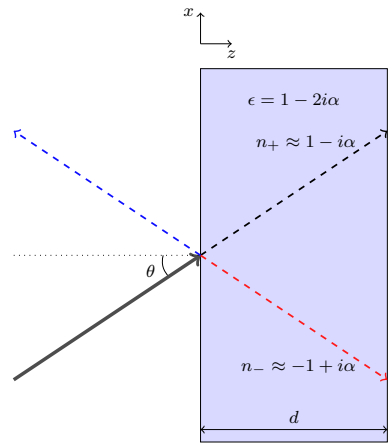}
  \caption{(Color online) Schematic view of a ray with oblique incidence to a nonmagnetic slab with $\epsilon=1-2i\alpha$.}
  \label{fig:obliqueschem}
\end{figure}

For a semi-infinite material the sign of $k_z'$ determines the reflection, refraction, and propagation of an incoming wave. In a slab the sign does not matter as the fields \eqref{slabrt} are invariant under $k_z'\to -k_z'$. This seems to be a contradiction: If a beam is incident to the slab at some angle of incidence $\theta$, the first-order transmitted beam must exit the slab {\it either} above or below the entrance point, see Fig. \ref{fig:obliqueschem}. A natural question arises: For a slab of thickness $d$, and $\epsilon=1-2i\alpha$, $0\leq\alpha\ll 1$ and $\mu=1$ at the excitation frequency, will the beam refract positively or negatively at the interfaces?
To investigate this problem, we consider active materials with $\epsilon=1-0.030i$ and $\mu=1$ at a normalized frequency $\omega=1$. We also assume that for any $d$, there is such a material that makes the slab electromagnetically stable (no global instabilities). This assumption will be justified below. Let the frequency-domain excitation be a beam at oblique incidence. The beam comprises a superposition of plane waves, whose amplitudes $A(k_x)$ are Gaussian distributed: $A(k_x)=1/\sqrt{2\pi\sigma^2}\exp[-(k_x-\bar k_x )^2/(2\sigma^2)]$ where $\sigma=0.032$, $\bar k_x=0.50$, and $0\leq k_x\leq 1$, normalizing the vacuum light velocity ($c=1$). To determine where the beam exits the slab, we consider the function $A(k_x)T(d,k_x)$, where $T=T(d,k_x)$ is the transmission coefficient \eqref{slabt}. By evaluating the inverse Fourier transform in $k_x$, we get the field $E(d,x)$ at the right-hand interface. Thus we can determine the correct solution (positively or negatively refracted beam) for different $d$. 
\begin{figure}[ht]
  \centering
  \includegraphics[width=9cm]{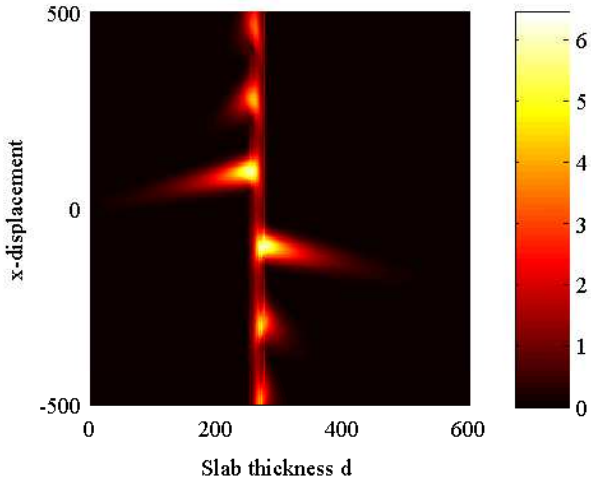}
  \caption{(Color online) The field amplitude $|E(d,x)|$ (arbitrary units) as a function of slab thickness $d$ and transversal coordinate $x$.}
  \label{fig:Txd}
\end{figure}
The field amplitude $|E(d,x)|$ is plotted in Fig. \ref{fig:Txd}. The beam undergoes multiple reflections in the slab; thus it exits the slab at several different locations along the $x$-axis. We observe that the beam refracts positively for $d\lessapprox 250$, and negatively for $d\gtrapprox 290$. For $250\lessapprox d\lessapprox 290$ the results indicate that there are no media with $\epsilon=1-0.03i$ that make the slab electromagnetically stable. 

\begin{figure}[ht]
  \centering
  \includegraphics[width=7cm]{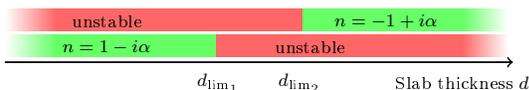}
  \caption{(Color online) Stability of nonmagnetic 1d slabs with $\epsilon=1-2i\alpha$.}
  \label{fig:schemview}
\end{figure}

The results certainly {\it do not} imply that for a given medium, the beam refracts positively or negatively dependent on the slab thickness. They rather imply that the media that make the slab stable for small thicknesses refract positively, while the media that make the slab stable for large thicknesses, refract negatively. This is indicated in Fig. \ref{fig:schemview}. The first category of media supports an amplifying, forward propagating wave, that is refracted positively. For this category, when $d\gtrapprox 250$ the round-trip gain becomes larger than the Fresnel losses, and the slab is unstable (global instability). The second category supports a backward propagating wave that is refracted negatively, and amplified as it propagates towards the left interface \cite{nistad07}. For this category, a global instability arises when $d\lessapprox 290$. When the slab thickness becomes large, the amplitude of the backward wave is small at the right-hand interface. This explains the decaying field amplitude as $d$ is increased above 290 in Fig. \ref{fig:Txd}. 

We now concretize the two categories of media by examples. Consider two nonmagnetic materials with $\epsilon_1(\omega)=1+\chi_g(\omega)$ and $\epsilon_2(\omega)=[1+\chi_r(\omega)]^2$, respectively, where $\chi_{r,g}(\omega)$ are Lorentzians in the form 
\begin{equation}\label{lorentzian}
\chi_{{r,g}}(\omega)=\frac{F_{r,g}\omega_{r,g}^2}{\omega_{r,g}^2-\omega^2-i\omega\Gamma_{r,g}}.
\end{equation}
With the parameters $F_g=-0.00016$, $\Gamma_g=0.005$, and $\omega_g=1$ for material 1, and $F_r=3.7$, $\Gamma_r=0.005$, and $\omega_r=0.6$ for material 2, these two materials have $\epsilon_{1,2}(\omega_g)=1-0.030i$. With the help of \eqref{eq:ndef}, we find the refractive indices $n_1(\omega_g)\approx1-0.015i$ and $n_2(\omega_g)\approx-1+0.015i$. These materials are both right-handed. They do not support absolute instabilities as $\epsilon_{1,2}(\omega)$ have no poles or zeros in the upper half-plane. Clearly, both of them show convective instabilities.

We should now investigate the stability as a function of $d$, when the slab is made of either of the two example materials. First we limit ourselves to a 1D slab. Active materials in 1D may for example be realized in the form of a transmission-line model with lumped circuit elements \cite{nistad07}. For 1D propagation, we take $k_x=k_y=0$ in \eqref{slabrt}, and determine numerically when there are no poles in the upper half-plane. We find that for $d<d_{\text{lim}_1}\approx 300$ the material 1 slab is stable, while for $d> d_{\text{lim}_2}\approx 320$ the material 2 slab is stable.

In the 2D case stability requires that there are no poles in the upper half-plane of \eqref{slabrt} for any $k_x$. For material 2, it turns out that the stability limit remains $d_{\text{lim}_2}\approx 320$. On the other hand, for material 1 the stability limit turns out to be zero ($d_{\text{lim}_1}=0$). By reducing the slab thickness, \eqref{slabrt} may not have poles for a fixed angle of incidence, but there will always be poles for a sufficiently large angle of incidence. This is due to the fact that as the angle of incidence increases towards $\pi/2$, the wave propagates longer between two reflections. Thus, during one round-trip, the wave picks up more gain compared to the losses associated with the two reflections. By limiting the transversal dimensions of the slab, and limiting the reflections at the upper and lower boundaries (e.g. using anti-reflection coatings or absorbing layers), this 2D instability can be eliminated. 

As we have seen, the fields of a slab are invariant to the sign of the refractive index. This is also the case for several other propagation problems such as plane wave scattering on a sphere or wave propagation in waveguides \cite{Lakhtakia:07oe}. However, the fact that the sign is invariant does not necessarily mean that it is irrelevant. The refractive index indicates the direction of the wave (``forward'' or ``backward'') before multiple reflections dominate the picture; it gives you information on whether the wave will refract positively or negatively at a boundary etc. Thus one can determine several properties of a given system by evaluating \eqref{eq:ndef}, without solving Maxwell's equations in detail. Nevertheless, for active materials it is important to be aware of possible instabilities, which may make frequency-domain concepts useless for real frequencies. The only way to fully understand whether a given shape of an active material is stable or not, is to solve Maxwell's equations with appropriate boundary conditions.

\section{Limitations from causality}\label{limitations}
Passive media, i.e., media in thermodynamic equilibrium in the absence of the variable field \cite{landau_lifshitz_edcm}, satisfy 
\begin{equation}
\im \epsilon(\omega)>0 \text{ and } \im \mu(\omega)>0 \text{ for } \omega>0 \label{loss}
\end{equation}
in addition to \eqref{KK}. This leads to fundamental limitations for left-handed media; for example, there is a lower bound for the dispersion associated with transparent, left-handed media, and there is a lower bound for the loss associated with nondispersive, left-handed media \cite{landau_lifshitz_edcm, smith_kroll, SS06}. 

Active media do not have the limitation \eqref{loss}, i.e., $\im\epsilon(\omega)$ and $\im\mu(\omega)$ may take any value. It is interesting to see if there are still any limitations resulting from symmetry \eqref{permittivitysym} and causality in the sense \eqref{KK}. It turns out that, in fact, {\it on a finite bandwidth, there are no fundamental limitations.} For example, $\epsilon(\omega)=\mu(\omega)=-1$  (lossless left-handedness) can be approached on a nonzero, finite bandwidth. This result follows directly from a standard result of the theory of Hardy spaces: A function $\epsilon(\omega)$ satisfying \eqref{permittivitysym} and \eqref{KK} can approximate any square integrable function $f(\omega)$ on a finite bandwidth \footnote{In mathematical terms, a function $h\in H^2$ satisfying $h(-\omega^*)=h^*(\omega)$ can approximate any function $f\in L^2(\omega_1,\omega_2)$, as precisely as desired in the corresponding metric. Here $H^2$ denotes the Hardy space of the upper half-plane, and $0\leq\omega_1<\omega_2<\infty$. See e.g. Ref. \cite{KrNu2}.}. The approximation may be achieved with any precision; in the Kre{\u\i}n--Nudel$'$man case \cite{KrNu2} at the expense of the norm of $\epsilon(\omega)-1$ outside the bandwidth of interest. Of course, a large norm outside the relevant bandwidth may imply difficulties of realization. Nevertheless, contrarily to claims elsewhere \cite{Stockman06}, we note that causality does not prohibit e.g. $\epsilon(\omega)=\mu(\omega)\approx -1$ with any precision, even in a finite bandwidth.

The possibility of approximating any desired behavior in a limited bandwidth may seem useless unless the resulting medium is free from absolute instabilities. With a straightforward approximation using e.g. Kre{\u\i}n--Nudel$'$man, the resulting permittivity may not be zero-free in the upper half-plane. A possible remedy is to approximate $\log f(\omega)$ with a function $g(\omega)$ using Kre{\u\i}n--Nudel$'$man, and setting $\epsilon(\omega)=\exp[g(\omega)]$. With this approach, it is necessary to assume that $\log f(\omega)$ is square integrable; in other words, $f(\omega)$ is not allowed to be zero on an interval of nonzero measure. This procedure clearly gives an analytic and zero-free function $\epsilon(\omega)$ in the upper half-plane. Square integrability of $\epsilon(\omega)-1$ (and Kramers--Kronig relations) are satisfied provided $\log f(\omega)$ is Lipschitz continuous and approaches zero at the end-points of the interval \cite{skaar01}. (The latter condition is not a further constraint as it may be fullfilled by extending the original interval.)
 
As an example of gain compensation of the losses associated with a left-handed resonance, consider the causal medium $\epsilon(\omega)=1+\chi_r(\omega)+\chi_g(\omega)$ and $\mu(\omega)=1+\chi_r(\omega)$, where $\chi_{r,g}(\omega)$ are the Lorentzians defined in \eqref{lorentzian}. Taking $\omega_r=1$, $\omega_g=1.058$, $F_r=0.25$, $F_g=-0.0034$, $\Gamma_r=0.005$, and $\Gamma_g=0.02$, we find that $n(\omega)=-1$ and $d\im n^2(\omega)/d\omega=0$ for $\omega=1.06$.
\begin{figure}[ht]
  \centering
  \includegraphics{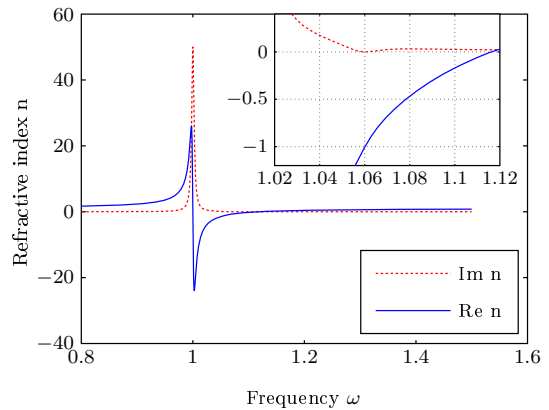}
  \caption{(Color online) Refractive index for the left-handed medium with gain compensation. At $\omega=1.06$ we have lossless negative refraction, $n=-1$ and $d\im n^2(\omega)/d\omega=0$.}
  \label{fig:stockman}
\end{figure}
This medium has $\im\epsilon(\omega)<0$ in the bandwidth $[1.05,1.07]$; thus it is net active there. This does not imply that the system is unstable unless the medium is infinite or put in a resonator configuration; examples of such stable systems include fiber optic amplifiers. As argued in Section \ref{instabilities}, the absence of absolute instabilities of this medium is guaranteed by the fact that $\epsilon(\omega)$ and $\mu(\omega)$ do not contain zeros or poles in the upper half-plane. We note that the gain compensation due to $\chi_g(\omega)$ has completely removed the loss which would have been present in the absence of this gain.

Lossless left-handedness in a finite bandwidth $(\omega_1,\omega_2)$ (with $0<\omega_1<\omega_2<\infty$) may in fact be obtained with a rather general class of susceptibility functions, even if we restrict ourselves to passive media. Pick any real, square integrable function $f(\omega)$, with $f(\omega)=0$ in $(\omega_1,\omega_2)$. Set $\im\epsilon(\omega)=\im\mu(\omega)=f(\omega)$. Thus, by tailoring $f(\omega)$ this medium is defined lossless at and near $\omega$, and may be defined lossy otherwise. Identifying $\re\epsilon(\omega)=\re\mu(\omega)$ with the Kramers--Kronig relation \eqref{KK2}, the resulting medium is causal by definition. From
\begin{eqnarray}
&&\re\epsilon(\omega)=\re\mu(\omega) \\
&&=1-\int_0^{\omega_1}\frac{2\omega'f(\omega')d\omega'}{\omega^2-\omega^{'2}}+\int_{\omega_2}^{\infty}\frac{2\omega'f(\omega')d\omega'}{\omega^{'2}-\omega^2},\nonumber
\end{eqnarray}
we observe that provided $f(\omega)$ is positive and sufficiently large below $\omega_1$, the medium will be left-handed. In fact, one can in principle construct causal, passive, negative index media with arbitrarily low (maximal) loss (see Appendix A).

\section{Bounds for the gain}\label{boundsgain}
The permittivity of an inverted Lorentzian medium, $\epsilon_1(\omega)=1+\chi_g(\omega)$, has a zero in the upper half-plane when $F_g<-1$. Thus a sufficiently strongly pumped medium seems have an absolute instability. It is natural to ask whether there is a general, upper bound on the gain for media without absolute instabilities. It turns out that it is not so. In fact, the second material in the previous section, $\epsilon_2(\omega)=[1+\chi_r(\omega)]^2$ and $\mu_2(\omega)=1$, does not have absolute instabilities no matter how large $F_r$ and therefore the maximum gain $\max_{\omega>0}[-\im\epsilon_2(\omega)]$ are. This is realized from the fact that $1+\chi_r(\omega)$ is a valid permittivity function of a passive medium, and does not have zeros in the upper half-plane.

\begin{figure}[ht] 
  \centering
  \includegraphics{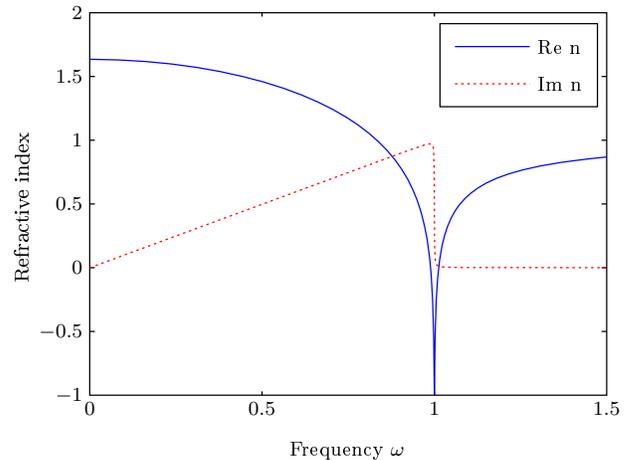}
  \caption{(Color online) Refractive index of a right-handed negative index material with low maximum gain. The required gain to obtain $\re n(\omega)=-1$ at some frequency can be made arbitrarily low by letting the edge of $\im n(\omega)$ be sufficiently steep. The refractive index is $n(\omega)=1+u(\omega)+iv(\omega)$, where $u(\omega)+iv(\omega)$ is the superposition \eqref{lorentzsuperposition} of Lorentzians. The parameters used are $\omega_1-\Delta\omega=1$ and $\Gamma=0.001$.}
  \label{fig:lowgain}
\end{figure}

Similarly, it is interesting to investigate if there is a least maximum gain to obtain a nonmagnetic (right-handed) negative index medium. A problem with the nonmagnetic negative index media that have been suggested so far \cite{chen05,nistad07}, is that they require very high gain in some spectral areas adjacent to the working frequency. This may imply difficulties for realizations, potential instability problems due to imperfections etc. We will now prove that it is indeed possible to reduce this gain without destroying the right-handed, negative index behavior. In fact, there exist nonmagnetic, causal media with arbitrarily low maximum gain, which refract negatively. To see this, we employ the result of Appendix \ref{maxloss}, which states that there are causal, passive, left-handed media with arbitrarily low maximum loss. Examples of such passive media are those given by $\epsilon(\omega)=\mu(\omega)=1+u(\omega)+iv(\omega)$, where the required properties of the susceptibility $u(\omega)+iv(\omega)$ are discussed in the Appendix. Letting our active medium instead have $\epsilon(\omega)=[1+u(\omega)+iv(\omega)]^2$ and $\mu(\omega)=1$, where $u(\omega)+iv(\omega)$ is the susceptibility of the passive medium in the Appendix, $n(\omega)$ must be identical to the refractive index of that passive medium. From the fact that the passive medium can have arbitrarily low maximum loss, we realize that the maximum gain of the present medium can be arbitrarily low. Thus, we have obtained a right-handed, negative index medium with arbitrarily low maximum gain. In Fig. \ref{fig:lowgain} the refractive index of such a medium is indicated, giving a rough guide to the realization of such media: The permittivity may e.g. be taken to be the square of a superposition of narrow Lorentzians. While a superposition of Lorentzians might be easy to realize approximately, it is not necessarily straightforward to achieve the square operation.

For such constructions from passive medium counterparts, \eqref{lowerbound} is valid for any refractive index functions that satisfy the Kramers--Kronig relations in the usual form \eqref{KK}. In other words, negative refraction requires either (i) large imaginary part below the working frequency, or (ii) exponentially steep variation immediately below the working frequency, or (iii) singularities at real frequencies.

For arbitrary active media that satisfy the Kramers--Kronig relactions in the usual form, inequality \eqref{lowerbound} remains valid if there is no gain above the observation frequency. If there is gain above the observation frequency $\omega_1$, it becomes
\begin{equation}
\label{boundgainloss}
u(\omega_1)\geq -\frac{2v_{\max}}{\pi}\left(\ln\frac{\omega_{\max}}{v_{\max}\Delta\omega}
+1\right),
\end{equation}
where $\omega_{\max}$ is the maximum frequency with gain, $1/\Delta\omega$ is the maximum steepness of $v(\omega)$, and $v_{\max}\equiv\max|\im n(\omega)|$. The proof of \eqref{boundgainloss} goes similarly to that of \eqref{lowerbound} in the Appendix. This shows that there is a trade-off between the maximum gain or loss, and the steepness.

There is however a possibility that negative refraction can be obtained with a nonmagnetic medium with large maximum loss (in accordance with \eqref{boundgainloss}) but small maximum gain. By letting $\epsilon(\omega)=1+\chi_r(\omega)+\chi_g(\omega)$, where $\Gamma_r=\Gamma_g$ is fixed, it is indeed possible to obtain a refractive index with $\re n<0$ in some spectral region, while $|F_g|$ is arbitrarily low. However, in the limit $F_g\to 0$, the figure of merit $-\re n/\im n$ tends to zero; thus these media may not be very useful.

\section{Conclusions}
Several questions related to active materials have been addressed. We have shown that Kramers--Kronig relations (for real frequencies) cannot always be established for causal, active media, due to the possibility of absolute instabilities. For active media that satify the Kramers--Kronig relations, causality imposes no fundamental limits to the dispersion and loss. 

Furthermore, we have considered the direction of the wave vector, and emphasized that if possible, the wave vector and the refractive index must be chosen analytic in the upper half-plane of complex frequency, and such that $n\to +1$, $k_z\to+\omega/c$ as $\omega\to\infty$. If the wave vector or refractive index have branch points or singularities in the upper half plane, the respective functions do not have meaning for real frequencies. Such nonanalytic points mean absolute instabilities, which are fundamentally different to the convective instabilities of conventional gain media, and global instabilities associated with bounded systems. 

Materials with $\epsilon=1-2i\alpha$ ($0<\alpha\ll 1$) and $\mu=1$ at a single frequency refract either positively or negatively. The materials that make a slab stable for small thicknesses, refract positively, while the materials that make the slab stable for large thicknesses, refract negatively. For a fixed thickness, at most one of the two categories of media makes the slab stable.

Finally we have argued that there are absolutely stable media with arbitrarily large gain. Moreover, we have proved that there are nonmagnetic negative index media with arbitrarily low maximum gain, giving directions to realizations. The proof is based on the fact that there are passive, left-handed media with arbitrarily low loss.

\appendix
\section{Passive left-handed media with arbitrarily low maximum loss}\label{maxloss}
\begin{figure}[ht]
  \centering
  \includegraphics{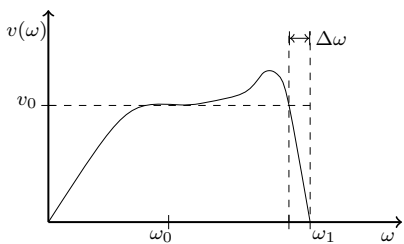}
  \caption{The function $v(\omega)$ has support below $\omega_1$. For $\omega_0<\omega<\omega_1-\Delta\omega$, we assume $v(\omega)\geq v_0$.}
  \label{fig:lowbound}
\end{figure}

Here we will show that there exist causal, passive, left-handed media with arbitrarily low maximum loss. The proof is by construction. Let the permittivity be written $\epsilon(\omega)=1+u(\omega)+iv(\omega)$, where $1+u(\omega)$ and $v(\omega)$ are the real and imaginary parts, respectively. Passitivity means that $v(\omega)>0$ for $\omega>0$. Since $v(\omega)=0$ can be approached, we allow ourselves to put $v(\omega)=0$ as well; adding a small, slowly varying function to $v(\omega)$ does not alter the argument below. Let $v(\omega)\geq v_0>0$ for $\omega_0<\omega<\omega_1-\Delta\omega$, and $v(\omega)=0$ for $\omega>\omega_1$, see Fig. \ref{fig:lowbound}. 
Estimating $u(\omega_1)$ by the Kramers--Kronig relation \eqref{KK2}, we find
\begin{eqnarray}
u(\omega_1)&&=-\frac{2}{\pi}\int_0^{\omega_1}\frac{v(\omega)\omega d\omega}{\omega_1^2-\omega^2}\nonumber\\
&&\leq -\frac{v_0}{\pi}\int_{\omega_0}^{\omega_1-\Delta\omega}\frac{2\omega d\omega}{\omega_1^2-\omega^2} \nonumber\\
&&= -\frac{v_0}{\pi}\left(\ln\frac{\omega_1-\omega_0}{\Delta\omega}
-\ln\frac{2\omega_1-\Delta\omega}{\omega_0+\omega_1}\right).
\label{upperbound}
\end{eqnarray}
Now we can choose any small $v_0$ and require the maximum of $v(\omega)$ to be less than, say $2v_0$. By having a sufficiently narrow transition region $\Delta\omega$, $u(\omega_1)$ gets less than $-1$. Setting $\mu(\omega)=\epsilon(\omega)=1+u(\omega)+iv(\omega)$ completes the proof.

Note that the required function can be approached by superpositions of several, narrow Lorentzians, with resonance frequencies equally spaced in the interval $(\omega_0,\omega_1-\Delta\omega)$. For example, in the limit of continuous varying resonance frequencies from $\omega_0=0$ to $\omega_1-\Delta\omega$, we obtain
\begin{eqnarray}
\label{lorentzsuperposition}
 u(\omega)+iv(\omega) &&\propto \int_0^{\omega_1-\Delta\omega} \frac{\omega_0^2d\omega_0}{\omega_0^2-\omega^2-i\omega\Gamma}\\
&&=\omega_1-\Delta\omega - s\arctan\left(\frac{\omega_1-\Delta\omega}{s}\right), \nonumber 
\end{eqnarray}
where $s=i\sqrt{\omega^2+i\omega\Gamma)}$. The required permittivity and permeability are obtained by choosing a sufficiently small $\Gamma$.

The steep edge in the transition band may imply that the medium is difficult to realize. Given a maximum loss $v_{\max}$ ($v(\omega)\leq v_{\max}$ for all $\omega$), one can prove that such steep edges is the only way to obtain left-handedness for media that satisfy Kramers--Kronig relations in the usual form \eqref{KK}. Indeed, for any square integrable function $v(\omega)\geq 0$ with limited steepness ($|dv(\omega)/d\omega|\leq 1/\Delta\omega$ for some $\Delta\omega$),
\begin{equation}
\label{lowerbound}
u(\omega_1)\geq -\frac{v_{\max}}{\pi}\left(\ln\frac{2\omega_1}{v_{\max}\Delta\omega}
+1\right).
\end{equation}
Here $\omega_1$ is an observation frequency. The inequality \eqref{lowerbound} is found by a similar argument as that of \eqref{upperbound}, considering the fact that the least possible $u(\omega_1)$ is obtained when $v(\omega)=v_{\max}$ for $0<\omega\leq\omega_1-\delta\omega$, and $v(\omega)$ decreases linearly to zero above $\omega_1-\delta\omega$. Here $\delta\omega$ is a positive parameter. Letting $1+u(\omega)+iv(\omega)$ be equal to $\epsilon(\omega)$, $\mu(\omega)$, or $n(\omega)$, the bound \eqref{lowerbound} applies to the permittivity, permeability, and refractive index of all passive media that satisfies the Kramers--Kronig relations in the conventional form \eqref{KK}. For media with singularities of $\epsilon(\omega)$ or $\mu(\omega)$ at real frequencies (e.g. an ideal plasma), \eqref{lowerbound} does not apply. We can conclude that left-handedness implies either (i) large loss below the working frequency, or (ii) exponentially steep variation immediately below the working frequency, or (iii) singularities at real frequencies. The trade-off between requirements (i) and (ii) is quantified by \eqref{lowerbound}. 
 
\def\cprime{$'$}

\end{document}